\documentclass[authoryear,preprint,11pt,a4paper]{elsarticle}

\usepackage{color,soul}

\usepackage{graphics}

\usepackage{epsfig}
\usepackage{rotating}

\usepackage{amssymb}

\usepackage{supertabular}

\usepackage{lscape}

\usepackage{fullpage}

\usepackage{lineno}

\journal{The Astrophysical Journal}

\begin{document}

\begin{frontmatter}

\title{Composition of Potentially Hazardous Asteroid (214869) 2007 PA8: An H Chondrite from the Outer Asteroid Belt}

\author[PSI]{Juan A. Sanchez\corref{cor1}\fnref{fn1}}
\ead{jsanchez@psi.edu}

\author[PSI]{Vishnu Reddy\fnref{fn1}}

\author[LPL]{Melissa Dykhuis}

\author[UO]{Sean Lindsay}

\author[PSI]{Lucille Le Corre\fnref{fn1}}

\address[PSI]{Planetary Science Institute, 1700 East Fort Lowell Road, Tucson, Arizona 85719, USA}

\address[LPL]{Lunar and Planetary Laboratory, University of Arizona, Tucson, AZ 85719, USA}

\address[UO]{Atmospheric, Oceanic, and Planetary Physics, University of Oxford, UK}

\fntext[fn1]{Visiting Astronomer at the Infrared Telescope Facility, which is operated by the University of Hawaii under Cooperative Agreement No. NNX-08AE38A with the National Aeronautics and Space Administration, 
Science Mission Directorate, Planetary Astronomy Program.}

\cortext[cor1]{Corresponding author at: Planetary Science Institute, 1700 East Fort Lowell Road, Tucson, Arizona 85719, USA.}

\begin{abstract}

Potentially hazardous asteroids (PHAs) represent a unique opportunity for physical characterization during their close approaches to Earth. The proximity of these asteroids makes them accessible for sample-return and 
manned missions, but could also represent a risk for life on Earth in the event of collision. Therefore, a detailed mineralogical analysis is a key component in planning future exploration missions and developing 
appropriate mitigation strategies. In this study we present near-infrared spectra ($\sim$ 0.7-2.55 $\mu$m) of PHA (214869) 2007 PA8 obtained with the NASA Infrared Telescope Facility during its close approach to Earth on 
November 2012. The mineralogical analysis of this asteroid revealed a surface composition consistent with H ordinary chondrites. In particular, we found that the olivine and pyroxene chemistries of 2007 PA8 are 
Fa$_{18}$(Fo$_{82})$ and Fs$_{16}$, respectively. The olivine-pyroxene abundance ratio was estimated to be 47\%. This low olivine abundance and the measured band parameters, close to the H4 and H5 chondrites, 
suggest that the parent body of 2007 PA8 experienced thermal metamorphism before being catastrophically disrupted. Based on the compositional affinity, proximity to the J5:2 resonance, and estimated flux of resonant 
objects we determined that the Koronis family is the most likely source region for 2007 PA8.

\end{abstract}


\end{frontmatter}


\section{Introduction}

Among the near-Earth asteroid (NEA) population, potentially hazardous asteroids (PHAs) are defined as objects having an absolute magnitude (H) $\leq$ 22.0 and a minimal orbital intersection distance (MOID) 
with Earth $\leq$ 0.05 AU. The study of these objects is considered of great importance because of their potential to make close approaches to the Earth and the risk they represent in the event of an impact. To date, 1540 
PHAs have been discovered\footnote{http://neo.jpl.nasa.gov/orbits/}, however only a small fraction of them have been compositionally characterized  
\citep[e.g.,][]{2009Icar..200..480B, 2012Icar..221..678R, 2012Icar..221.1177R, 2013Icar..225..131S}. This information is essential to assess the level of damage that they could cause in the eventuality of collision and to 
develop future mitigation strategies.

The PHA (214869) 2007 PA8 was discovered by the LINEAR survey on August 9$^{\rm{th}}$, 2007.  During its close approach to Earth in 2012 several teams were able to carry out observations using different 
techniques. Broadband photometric data (Bessel $BVRI$) obtained by \citet{2012ATel.4625....1H} showed that the average colors, $B-R$, $V-R$, and $R-I$, of this asteroid were consistent with an Xc-type (Bus Taxonomy, 
\citet{2002Icar..158..146B}) or a C-type classification (Tholen Taxonomy, \citet{1984PhDT.........3T}). \citet{2013DPS....4510109B} obtained radar images of 2007 PA8 and found that this object has an elongated shape with 
dimensions of 1.9x1.4x1.3 km, and a slow rotation period of 101.6 $\pm$ 2.0 h. Near infrared (NIR) spectral data (0.8-2.5 $\mu$m) acquired by \citet{2014A&A...567L...7N} with the Infrared Telescope Facility (IRTF) showed 
that the NIR spectrum of 2007 PA8 is similar to that of H chondrites. \citet{2015Icar..250..280F} also obtained NIR spectra of 2007 PA8 using the Telescopio Nazionale Galileo (TNG). These data were combined with visible 
spectra in order to extend the wavelength range (0.38-2.4 $\mu$m). From the mineralogical analysis they determined that this asteroid has a composition consistent with L chondrites, and that it probably originated in the 
Gefion family. 

In this paper we present new NIR spectra of 2007 PA8 obtained during its close approach to Earth in 2012. Our data have the advantage of having higher quality than data presented in previous studies, and a wavelength 
range that allows a more accurate mineralogical analysis of this asteroid. This will help to clarify the apparent discrepancies among the previous studies regarding the composition and possible 
origin of 2007 PA8. In section 2 we describe the observations and data reduction procedure. In section 3 we present the results of our mineralogical analysis and we compare them with previous studies. We also discuss the 
possible source region for 2007 PA8. Finally, in section 4 we summarize our main findings.

\section{Observations and data reduction}

NIR spectra of 2007 PA8 were obtained using the SpeX instrument \citep{2003PASP..115..362R} on the NASA IRTF on Mauna Kea, HawaiÕi on November 1, 2012. Weather conditions were 
photometric during the observing run with a seeing of $\sim$0.55" and relative humidity of $\sim$5\%. Asteroid spectra were acquired using SpeX in its low-resolution (R$\sim$150) prism 
mode with a 0.8" slit width, providing an effective wavelength range of $\sim$ 0.7-2.55 $\mu$m. Spectra were obtained using a nodding technique in which the asteroid is alternated between 
two different slit positions (A and B) following the sequence ABBA. During the observations the slit was oriented along the parallactic angle to minimize differential refraction at the shorter wavelength end.

We acquired twenty spectra of 2007 PA8 when the asteroid was 11.6 visual magnitude at a heliocentric distance r=1.03 AU, and a phase angle of $41.2^\mathrm{o}$. Each exposure was limited to 
120 seconds. In addition to 2007 PA8, G-type local extinction star HD24821 was observed before and after the asteroid observations in order to correct for telluric features. Twenty spectra of solar analog 
star SAO93936 were also obtained to correct for any spectral slope variation caused by the use of a non-solar extinction star.

\begin{figure*}[!ht]
\begin{center}
\includegraphics[height=10cm]{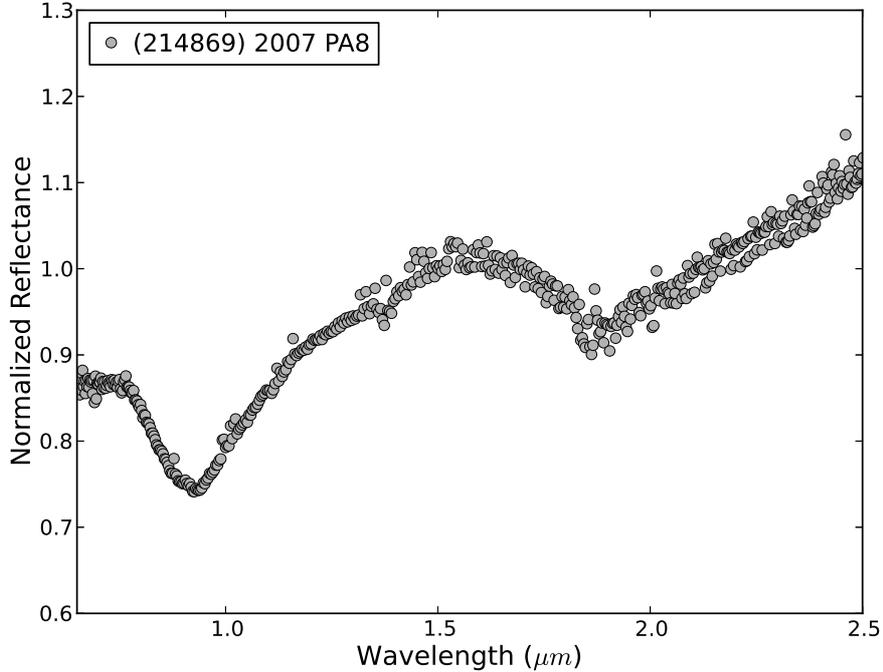}
\caption{\label{f:2007PA8spec} {\small NIR spectrum of asteroid (214869) 2007 PA8 obtained using the SpeX instrument on NASA IRTF. The spectrum was normalized to unity at 1.5 $\mu$m. The two absorption bands 
centered at $\sim$ 0.94 and 1.91 $\mu$m are characteristics of olivine-pyroxene assemblages.}}
\end{center}
\end{figure*}

NIR spectral data were reduced using the IDL-based software Spextool \citep{2004PASP..116..362C}. The steps followed in the reduction process include: 1) sky background removal by subtracting the 
A-B image pairs, 2) flat-fielding, 3) cosmic ray and spurious hit removals, 4) wavelength calibration, 5) division of asteroid spectra by the spectrum of the solar analog star, and 6) co-adding of individual spectra. The NIR 
spectrum of asteroid 2007 PA8 is shown in Figure \ref{f:2007PA8spec}. Diagnostic spectral band parameters including Band I and II centers, Band depths, and Band Area Ratio (BAR) were measured using a Python code 
following the protocols described in \citet{1986JGR....9111641C} and \cite{2002aste.conf..183G}. In our case we defined the unresolved red edge of the spectrum as 2.5 $\mu$m. The uncertainties associated to the band 
parameters are given by the standard deviation of the mean calculated from multiple measurements of each band parameter. The average surface temperature of the asteroid at the time of observation was calculated as in 
\citet{2009M&PS...44.1331B}. This value was used along with equation 7 from \cite{2012Icar..220..36S} to apply a temperature correction to the BAR. Spectral band parameters and their uncertainties are presented in 
Table \ref{t:Table1}.

\begin{table}[!ht]
\caption{\label{t:Table1} {\small Spectral band parameters, molar contents of fayalite (Fa) and ferrosilite (Fs), and olivine-pyroxene abundance ratio (ol/(ol+px)) for 2007 PA8.}}
\begin{center}
\begin{tabular}{cr}
\hline
Parameter&Value \\
\hline
Band I Center ($\mu$m)&0.935$\pm$0.005 \\
Band II Center ($\mu$m)&1.91$\pm$0.01 \\
Band I Depth (\%)&17.5$\pm$0.2 \\
Band II Depth (\%)&11.0$\pm$0.5 \\
Band Area Ratio (BAR)&1.08$\pm$0.05 \\
Temp. corrected BAR&1.06$\pm$0.05 \\
Olivine Composition (mol \%)&Fa$_{18.2\pm1.3}$ \\
Pyroxene Composition (mol \%)&Fs$_{16.1\pm1.4}$ \\
ol/(ol+px)&0.47$\pm$0.03 \\

\hline
\end{tabular}
\end{center}
\end{table}

\clearpage

\section{Results}

\subsection{Compositional analysis}

\citet{2014A&A...567L...7N} found that the closest taxonomic type to 2007 PA8 was the Sq in the Bus-DeMeo taxonomic system \citep{2009Icar..202..160D}. \citet{2015Icar..250..280F}, on the other hand, classified this 
object as a Q-type in the same taxonomic system. Using the online Bus-DeMeo taxonomy calculator (http://smass.mit.edu/busdemeoclass.html) we determined that 2007 PA8 has a 
PC1'= -0.4609 and a PC2'=0.2359, which places this asteroid in the Q-type region of the PC1' vs PC2' diagram of \citet{2009Icar..202..160D}, consistent with the results obtained by 
\citet{2015Icar..250..280F}. 

\begin{figure*}[!ht]
\begin{center}
\includegraphics[height=10cm]{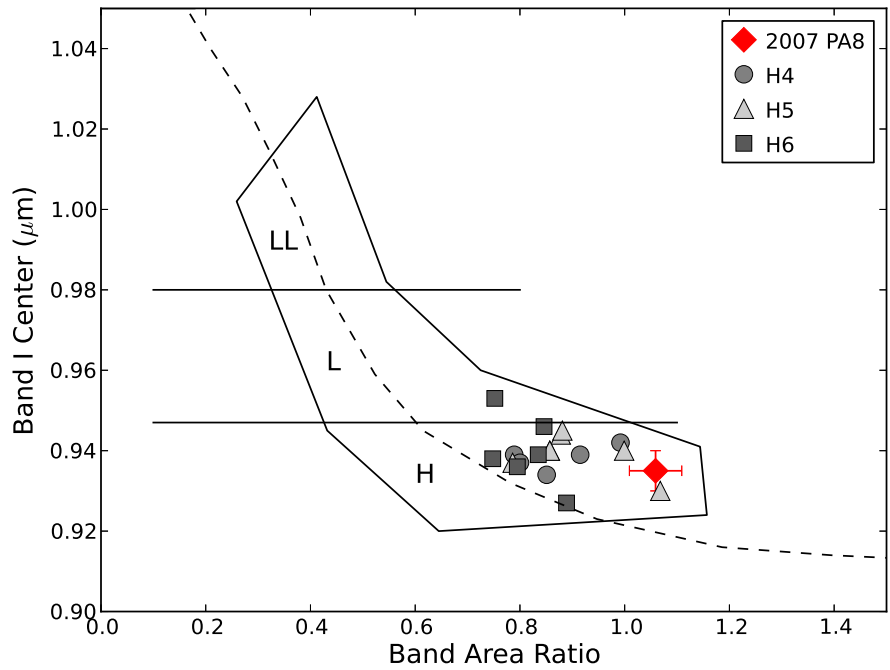}
\caption{\label{f:BICBAR} {\small Plot of the Band I center vs. BAR for 2007 PA8 (red diamond). Also shown, the values measured for H ordinary chondrites from \citet{2010Icar..208..789D}. The polygonal region 
corresponds to the S(IV) subgroup of \citet{1993Icar..106..573G}. The horizontal lines represent the approximate boundaries for ordinary chondrites found by \citet{2010Icar..208..789D}. The dashed line indicates the 
location of the olivine-orthopyroxene mixing line of \citet{1986JGR....9111641C}.}}
\end{center}
\end{figure*}

Taxonomic classification alone does not provide information about the surface composition of the body, therefore the next step in our analysis is to plot the measured band 
parameters in the Band I center vs. BAR diagram \citep{1993Icar..106..573G}. As can be seen in Figure \ref{f:BICBAR} the values measured for 2007 PA8 (red diamond) are located inside the polygonal region 
corresponding to the S(IV) subgroup of \citet{1993Icar..106..573G}, in the H ordinary chondrite zone. \citet{1998M&PS...33.1281G} noticed that within the petrologic subtypes of the H chondrites there is a shift 
from left to right in the Band I center vs. BAR diagram from H6 to H4 chondrites. In Figure \ref{f:BICBAR} we plotted the values measured by \citet{2010Icar..208..789D} for the equilibrated H chondrites (types 4-6). The 
distribution of the different petrologic subtypes is consistent with the findings of \citet{1998M&PS...33.1281G}, where the H6 are grouped to the left and the H4 and H5 tend to move to the right. Thus, the position of  2007 PA8 
in Figure \ref{f:BICBAR} could suggest an affinity with either H4 or H5 chondrites.

\begin{figure*}[!ht]
\begin{center}
\includegraphics[height=10cm]{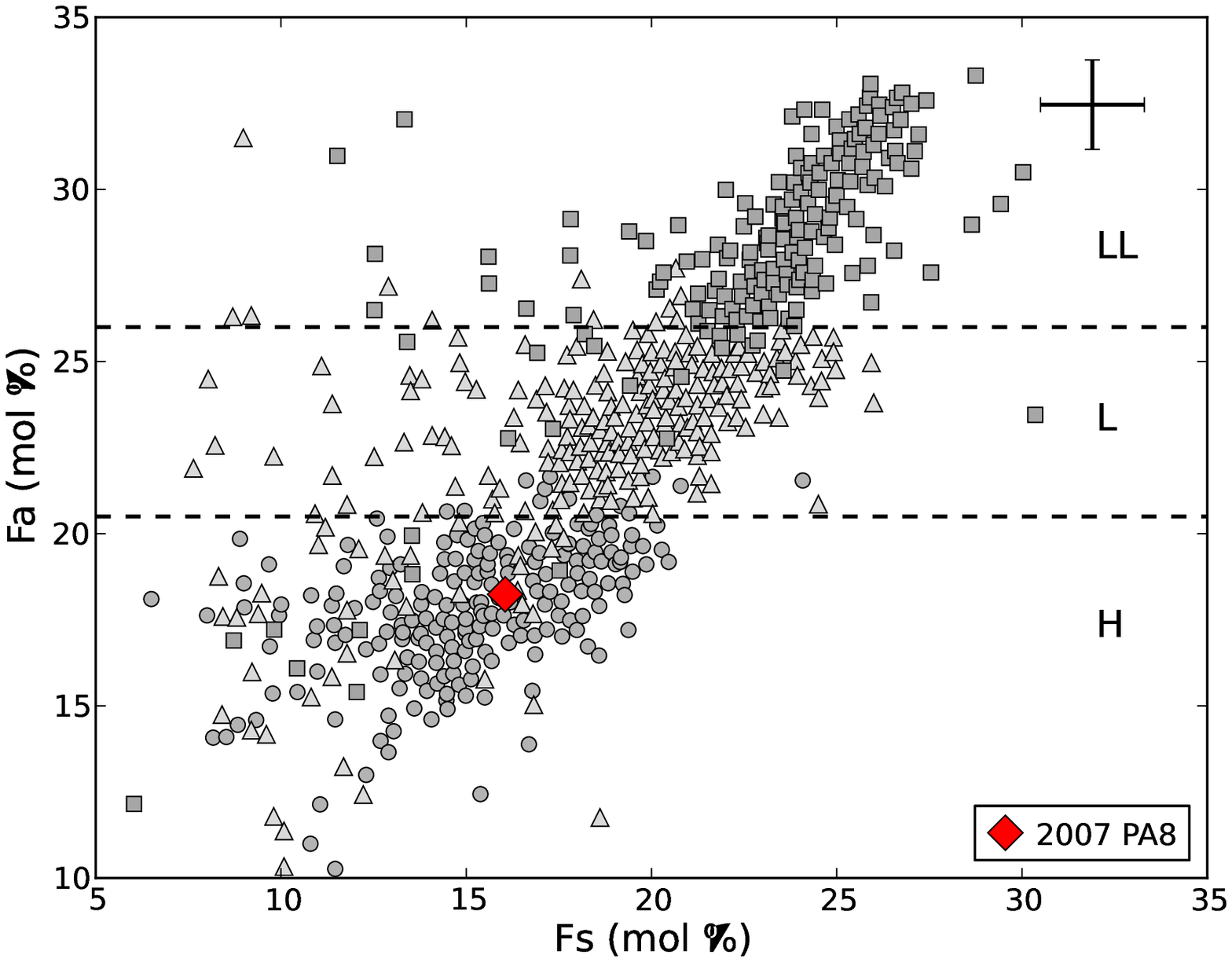}

\caption{\label{f:FaFs5} {\small Molar contents of fayalite (Fa) vs. ferrosilite (Fs) for 2007 PA8 (red diamond), along with the measured values for LL (squares), L (triangles), and 
H (circles) ordinary chondrites from \citet{2011Sci...333.1113N}. The error bars in the upper right corner correspond to the uncertainties derived 
by \citet{2010Icar..208..789D}, 1.3 mol\% for Fa, and 1.4 mol\% for Fs. Figure adapted from \citet{2011Sci...333.1113N}.}}

\end{center}
\end{figure*}

\begin{figure*}[!ht]
\begin{center}
\includegraphics[height=10cm]{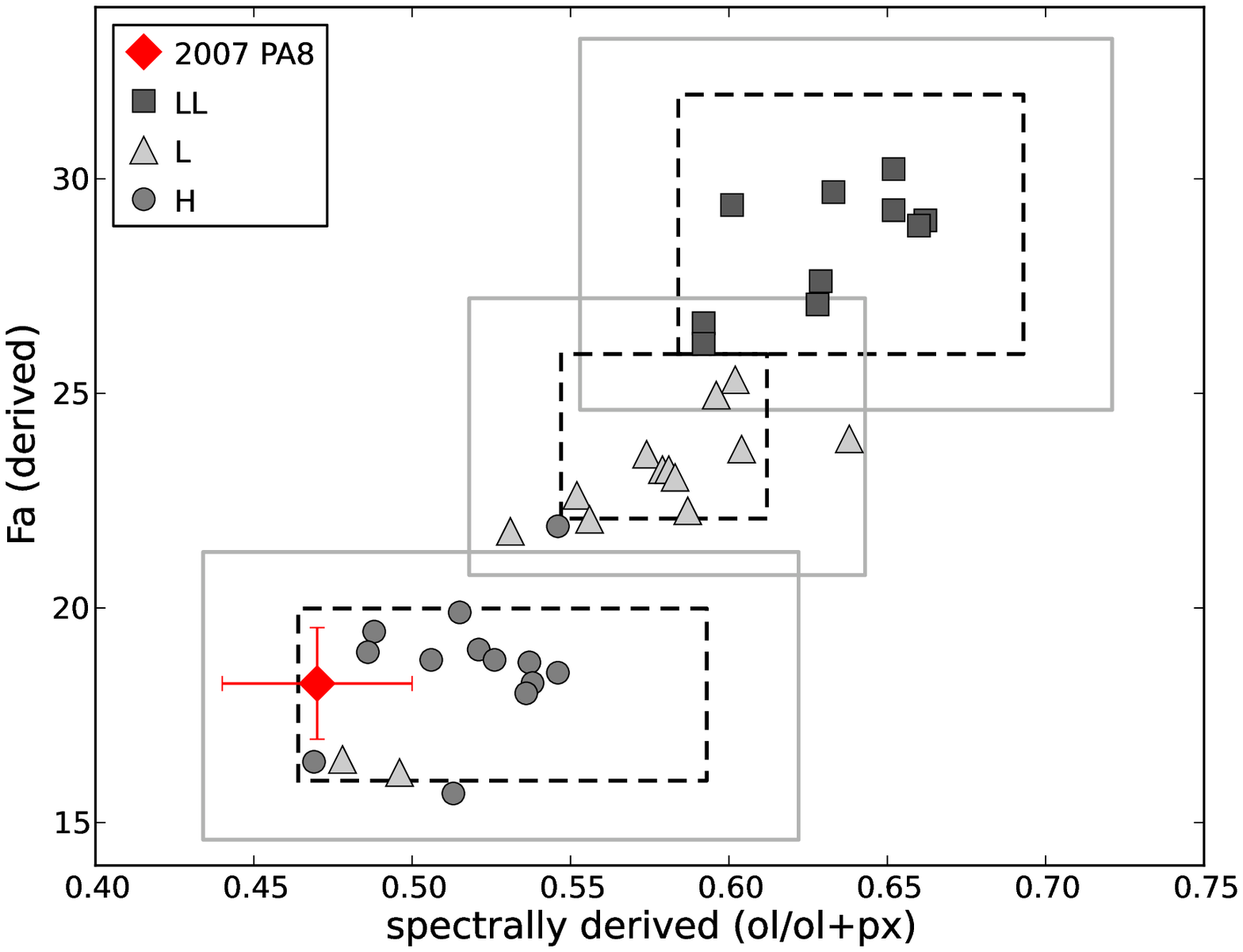}
\caption{\label{f:Faol} {\small Molar content of Fa vs. ol/(ol+px) ratio for 2007 PA8 (red diamond), along with the measured values for LL, L and H ordinary chondrites from 
\citet{2010Icar..208..789D}. Black dashed boxes represent the range of measured values for each ordinary chondrite subgroup. Gray solid boxes correspond to the 
uncertainties associated to the spectrally-derived values. Figure adapted from \citet{2010Icar..208..789D}.}}
\end{center}
\end{figure*}

\begin{figure*}[!ht]
\begin{center}
\includegraphics[height=10cm]{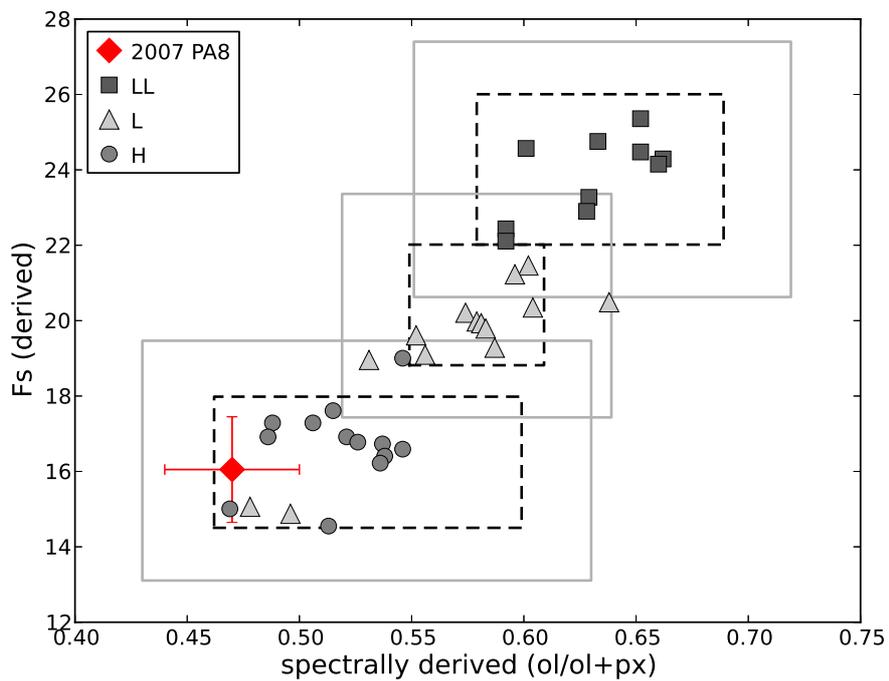}
\caption{\label{f:Fsol} {\small Molar content of Fs vs. ol/(ol+px) ratio for 2007 PA8 (red diamond), along with the measured values for LL, L and H ordinary chondrites from 
\citet{2010Icar..208..789D}. Black dashed boxes represent the range of measured values for each ordinary chondrite subgroup. Gray solid boxes correspond to the 
uncertainties associated to the spectrally-derived values. Figure adapted from \citet{2010Icar..208..789D}.}}
\end{center}
\end{figure*}

Since the Band I center and BAR of 2007 PA8 fall inside the ordinary chondrites region, we used the equations derived by \citet{2010Icar..208..789D} to determine the surface 
composition of the asteroid. We found that the olivine and pyroxene chemistries of 2007 PA8, given by the molar contents of fayalite (Fa) and ferrosilite (Fs) 
respectively, are Fa$_{18}$(Fo$_{82})$ and Fs$_{16}$. These values are consistent with those derived by \citet{2010Icar..208..789D} for H ordinary 
chondrites (Fa$_{16-20\pm 1.3}$ and Fs$_{14-18\pm 1.4}$). In Figure \ref{f:FaFs5} we plotted the molar content of Fa vs. molar content of Fs for 2007 PA8, along with measured values for LL, L, and 
H ordinary chondrites from \citet{2011Sci...333.1113N}. The olivine-pyroxene abundance ratio (ol/(ol+px)) was estimated in 0.47, again consistent with measured values for H chondrites (0.46-0.60$\pm 0.03$) obtained 
by \citet{2010Icar..208..789D}. Figures \ref{f:Faol} and \ref{f:Fsol} show the molar content of Fa and Fs for 2007 PA8 (red diamond) vs. ol/(ol+px) ratio. As can be seen in these figures the derived values for 2007 PA8 fall in 
the H ordinary chondrite region.

\clearpage

\subsection{Comparison with previous studies}

The results presented in the previous section seem to be consistent with the findings of \citet{2014A&A...567L...7N}, but inconsistent with the results obtained by 
\citet{2015Icar..250..280F}. The NIR spectrum (0.8-2.5 $\mu$m) of 2007 PA8 acquired by \citet{2014A&A...567L...7N} did not have a shorter wavelength rollover for the 
1-$\mu$m band. Because of that they were not able to measure the Band I center and BAR from their spectrum, preventing a detailed mineralogical analysis of the 
asteroid. However, using curve matching with laboratory spectra they found that the best fit for the spectrum of 2007 PA8 was represented by the H5 Cangas de Onis 
ordinary chondrite. Despite the fact that curve matching is a non-diagnostic technique, the results obtained by \citet{2014A&A...567L...7N} are in good agreement with 
our mineralogical analysis that shows that 2007 PA8 has a surface composition similar to H4-5 ordinary chondrites. 

The spectrum of 2007 PA8 obtained by \citet{2015Icar..250..280F} did have data shortward of 0.8 $\mu$m, covering the wavelength range of 
0.38-2.4 $\mu$m. The technique used by \citet{2015Icar..250..280F} to measure the band parameters was very similar to the one we used, however the values that they 
obtained show important differences compared to ours. They found a Band I center of 0.9578 $\mu$m, with a Band I depth of 16.5\%, and a Band II 
center of 1.958 $\mu$m, with a Band II depth of $\sim$ 4\%. From their measurements, \citet{2015Icar..250..280F} calculated a value for the BAR of 0.4. They used these values and the equations of 
 \citet{2010Icar..208..789D} to derive the surface composition of 2007 PA8, and found olivine and pyroxene chemistries of Fa$_{23.1\pm 0.8}$ and Fs$_{19.7\pm 0.6}$ respectively, 
 with an ol/(ol+px) of 0.63$\pm 0.02$. These values are consistent with those estimated for L chondrites. In contrast, we found that the Band I center is located at 0.935 $\mu$m and the 
 Band II center is at 1.91 $\mu$m, with Band I and II depths of 17.5 and 11.0\%, and a BAR of 1.06. These differences could be attributed to various factors, including: 
 surface heterogeneities, differences in the computer code used to measure the band parameters, the observational circumstances, the wavelength range covered, and the 
 quality of the data. 
 
 Surface heterogeneities caused by compositional variations were considered by \citet{2015Icar..250..280F} as a possible explanation for the spectral 
 differences observed between their spectrum and the one obtained by \citet{2014A&A...567L...7N}. In order to corroborate this hypothesis we would have to obtain rotationally resolved spectra of 
 2007 PA8, and look for possible hemispherical compositional variations. However, given the slow rotation period of this asteroid ($\sim$102 h) this would be very difficult to accomplish. Although 
 compositional variations can not be completely ruled out they are rarely seen among small objects like NEAs \citep[see discussion in][]{2015Icar..252..129R}, and therefore, we consider this possibility remote compared to 
 the other possible explanations (see below). 
  
 In order to rule out possible problems with our code we used two additional programs to measure the band parameters, the Spectral Analysis Routine for Asteroids (SARA), developed by \citet{2015Icar..247...53L}, which is 
 an IDL-based band parameter analysis routine, and the Modeling for Asteroid Spectra (M4AST) package developed by \citet{2012A&A...544A.130P}. This is an online tool\footnote{http://m4ast.imcce.fr} used for modeling 
 visible and NIR spectra of asteroids, and it is the same tool used by \citet{2014A&A...567L...7N}. The results obtained in both cases are consistent with those obtained with our Python code.
  
 Regarding the observational circumstances we noticed that \citet{2015Icar..250..280F} observed 2007 PA8 at a phase angle of $\sim11^\mathrm{o}$, while our 
 observations were obtained at a phase angle of $41.2^\mathrm{o}$. An increase in phase angle can produce variations in the strength of the absorption bands and 
 increase of the spectral slope \citep{2012Icar..220..36S}. This could explain the deeper absorption bands and steeper spectral slope exhibited by our spectrum 
 compared to the spectrum obtained by \citet{2015Icar..250..280F}. However, as explained by \citet{2012Icar..220..36S}, an increase in phase angle will not have a 
 significant effect on the Band I center and BAR, and therefore the effect on the mineralogical analysis will be negligible.

 The wavelength range covered, and the quality of the data are the most likely cause for the differences seen in the band parameters and the derived surface composition. 
 \citet{2015Icar..250..280F} used two different instruments to acquire the visible and NIR spectra of 2007 PA8. The visible spectrum was obtained using the Dolores 
 instrument, which is equipped with two low resolution grisms that combined cover the 0.38-0.95 $\mu$m range. The NIR spectrum, on the other hand, was obtained 
 using the near infrared camera and spectrometer (NICS), which covers the 0.85-2.4 $\mu$m range. The visible and NIR spectra were combined by overlapping the 
 spectra at the common wavelengths. Since the Band I center is measured near the point where the visible and NIR spectra are merged, it is possible that even a small 
 shift could have influenced this parameter, which in turns will affect the calculation of the olivine and pyroxene chemistries. The spectrum obtained by 
 \citet{2015Icar..250..280F} is also truncated at 2.4 $\mu$m, and as a result, measurements of the Band II depth and Band II area will be underestimated. This is a problem 
 acknowledged by \citet{2015Icar..250..280F}, which will affect the Band Area Ratio and hence the estimation of the olivine-pyroxene abundance ratio. In addition, we also note that in the 
 spectrum obtained by \citet{2015Icar..250..280F} data between 1.15-1.2 $\mu$m, 1.3-1.5 $\mu$m and 1.75-2.0 $\mu$m were deleted due to incomplete correction of the telluric bands. 
 Noisy data or incomplete data is one of the factors that most affect the spectral band parameters.

\subsection{Source region and possible parent bodies}

H ordinary chondrites have been traditionally linked to main belt asteroid (6) Hebe, based on their spectral similarities and compositional affinity \citep{1998M&PS...33.1281G}. In addition, the 
proximity of (6) Hebe (located at 2.426 AU) to the 3:1 mean motion resonance provides a viable route for the delivery of fragments of this object to the near-Earth space. The orbit of 2007 
PA8 ($a$ = 2.82 AU, $e$ = 0.66, $i = 1.98^\mathrm{o}$), however suggests that this asteroid originated in the outer part of the main belt. This was confirmed by \citet{2014A&A...567L...7N}, who 
studied the dynamical evolution of 2007 PA8 by generating a population of 1275 clones. These clones were then integrated backward in time for 200,000 years using a dynamical model. The results obtained by 
\citet{2014A&A...567L...7N} showed that after being injected into the J5:2 mean motion resonance (at $\sim$ 2.82 AU) the orbit of 2007 PA8 evolved from the main belt to an Earth-crossing orbit on a time scale of 10$^{5}$ 
years. 

\begin{figure*}[!ht]
\begin{center}
\includegraphics[height=11cm]{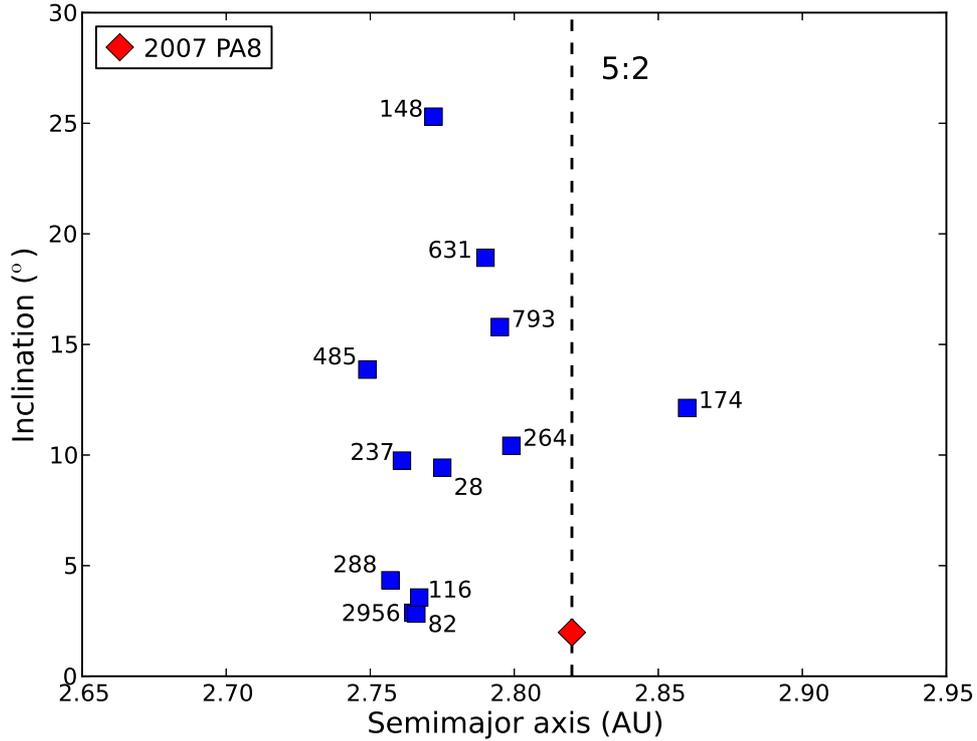}
\caption{\label{f:Hasteroids3} {\small Inclination vs. semimajor axis for 2007 PA8 (red diamond), and a group of main belt asteroids with H chondrite-like compositions (blue squares) identified by 
\citet{2014ApJ...791..120V}. The number of each asteroid is indicated. The location of the J5:2 mean motion resonance is represented by a vertical dashed line.}}
\end{center}
\end{figure*}

Further evidence supporting the outer main belt as a source region for H chondrite parent bodies comes from a recent study carried out by \citet{2014ApJ...791..120V}. They obtained NIR spectra of 83 main belt 
S-type asteroids and six asteroid families, and found that several objects with orbits near the J5:2 resonance have surface compositions consistent with H chondrites. In Figure \ref{f:Hasteroids3} we plot inclination vs. 
semimajor axis for the asteroids with H chondrite-like compositions with the closest orbits to the J5:2 resonance. \citet{2014ApJ...791..120V} used a radiative transfer model \citep{1999Icar..137..235S} to estimate the 
ol/(ol+px) ratio of these asteroids. For the objects depicted in Figure \ref{f:Hasteroids3} the olivine abundance ranges from 51 to 63\%, which is higher than the calculated value for 2007 PA8 (47\%). It is important to point out, 
however, that the technique used by \citet{2014ApJ...791..120V} to calculate the olivine abundance of the asteroids is different from the one we employ. We noticed that in \citet{2008Natur.454..858V} the use of the radiative 
transfer model to estimate the ol/(ol+px) ratio in ordinary chondrites gave an average value of $\sim$ 59\% for H chondrites. This value is higher than the average value (52\%) derived by \citet{2010Icar..208..789D} using 
laboratory spectral calibrations (the technique used in the present study). Therefore, in order to establish a more accurate comparison, we applied a correction factor of 7\% to the values derived by 
\citet{2014ApJ...791..120V}. This results in asteroids having ol/(ol+px) ratios that range from 44 to 56\%, which span the value derived for 2007 PA8. 

The asteroid families studied by \citet{2014ApJ...791..120V} included: Agnia, Merxia, Koronis, Gefion, Eunomia, and Flora (for comparison with \citet{2015aste.conf..N}, these families have Family Identification 
Numbers, or FINs, of 514, 513, 605, 516, 502, and 402, respectively). The first three families were found to have olivine abundances consistent with H chondrites, $\sim$43-53\% for Agnia and Merxia, and $\sim$47-54\% for 
the Koronis family (in both cases after applying the correction factor of 7\%), while Gefion, Eunomia, and Flora show surface properties consistent with L and LL chondrites. Hence, Agnia, Merxia and Koronis are the 
strongest candidates for the source region of 2007 PA8.

Apart from the compositional similarities between 2007 PA8 and the Agnia, Merxia and Koronis family, another possible link between this asteroid and the families is the apparent thermal metamorphism experienced by 
2007 PA8.  \citet{2014ApJ...791..120V} determined that unequilibrated H chondrites (type 3.0 to 3.4) have ol/(ol+px) ratios $>$ 65\%, while equilibrated or thermally metamorphosed H chondrites (type 3.6 to 6) have 
ol/(ol+px) ratios $<$ 65\%. In the case of the Agnia, Merxia, and Koronis family, \citet{2014ApJ...791..120V} found that they all have olivine abundances $<$ 65\% (58\% if we apply the correction factor of 7\%), i.e., 
compatible with equilibrated H chondrites. Thus, the olivine abundance of 2007 PA8 (47\%) and its location in the Band I center vs. BAR diagram, close to the H4 and H5 chondrites (Figure \ref{f:BICBAR}), would support the 
scenario in which this object originated in the interior of the thermally metamorphosed parent body of one of these families. \citet{2013GeCoA.119..302M} derived depth ranges for the different petrologic 
types of H chondrites assuming that their parent body had an "onion shell" structure (i.e., material of increasing petrologic type originates from deeper layers). They determined that the most metamorphosed 
material (type 6) was located in the central part of the body, up to a radius r=0.64, while types 5 and 4 should have occupied concentric shells from r = 0.64 to r = 0.98 in their parent body. 
Therefore, the petrologic type inferred for 2007 PA8 indicates that this object originated in the middle-upper layers of its parent body, where the boundary between petrologic types H4/5 
reached temperatures of 1000 K \citep{2013GeCoA.119..302M}.

\begin{figure*}[!ht]
\begin{center}
\includegraphics[height=11cm]{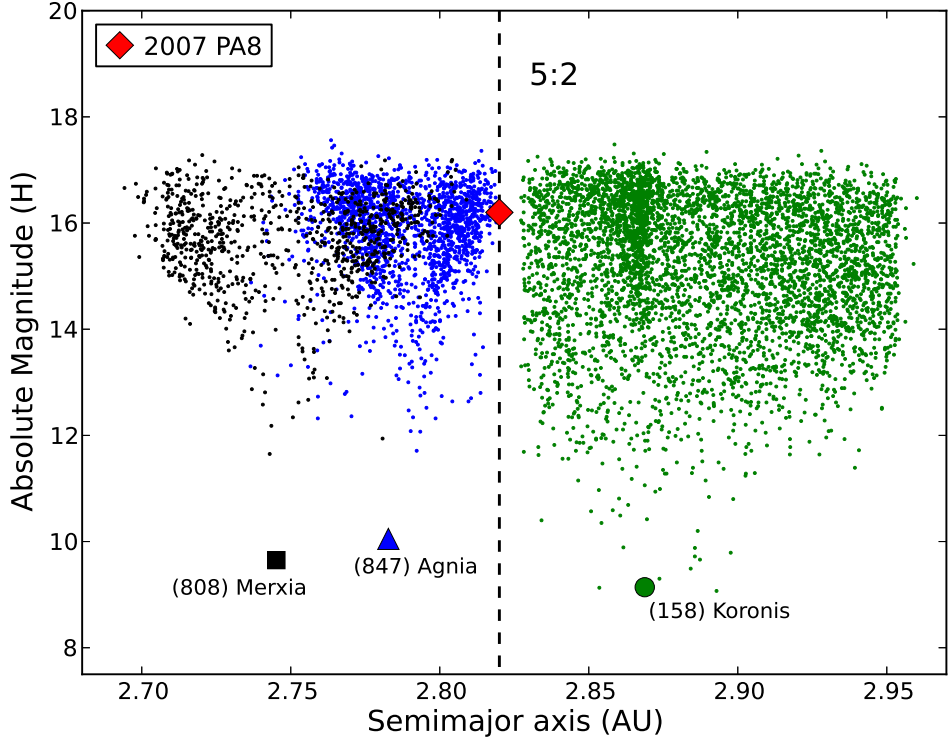}
\caption{\label{f:Ha} {\small Absolute magnitude vs. semimajor axis for 2007 PA8 (red diamond) and asteroid families Merxia (black), Agnia (blue) and Koronis (green) from \citet{2012PDSS..189.....N}. Also 
shown the location of asteroids (158) Koronis (green circle), (808) Merxia (black square) and (847) Agnia (blue triangle). The location of the J5:2 mean motion resonance is represented by a vertical dashed line. The three 
families display "V'" shapes indicating Yarkovsky-driven semimajor axis evolution (although the V for the older Koronis family is heavily truncated at the J5:2 resonance). The expected flux of PA8-sized objects into the J5:2 
is much higher from the Koronis family than from the Merxia and Agnia families.}}
\end{center}
\end{figure*}

Additional constraints on the source region for 2007 PA8 come from the expected contribution rates of resonant objects in the size range of 2007 PA8 (1.3 to 1.9 km) from the dynamical families with H chondrite-like 
compositions bordering the J5:2 resonance. In order to estimate the expected contributions we plot the semimajor axis and absolute magnitude distribution of the Agnia, Merxia and Koronis family in Figure 
\ref{f:Ha}. The three families show distinct ``V'' shapes in this parameter space, indicative of their size-dependent semimajor axis dispersion due to the Yarkovsky effect \citep{2006Icar..182..118V}.

\begin{figure*}[!ht]
\begin{center}
\includegraphics[height=11cm]{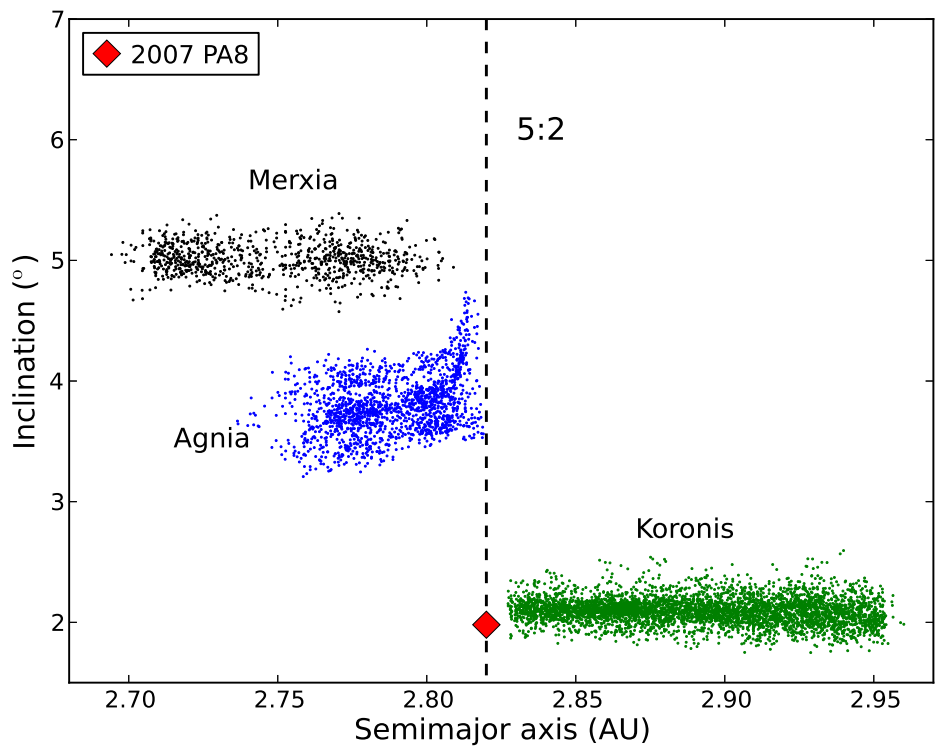}
\caption{\label{f:ia} {\small Inclination vs. semimajor axis for 2007 PA8 (red diamond), and asteroid families Merxia (black), Agnia (blue), and Koronis (green) from \citet{2012PDSS..189.....N}. The location of the J5:2 
mean motion resonance is represented by a vertical dashed line.}}
\end{center}
\end{figure*}

The wider V shape of the Koronis family indicates the family's older age (1-2 By), and the truncation of the V at the J5:2 resonance implies that a significant number of Koronis family members, including those in the same 
size range as 2007 PA8, have drifted into the resonance and been removed from the region. By contrast, the Merxia family is not currently expected to produce resonant PA8-sized objects, as its Yarkovsky V is not dispersed 
enough yet (i.e., the family is too young) for objects in this size range to reach the J5:2. The Yarkovsky V for the Agnia family is even less dispersed, indicating an even younger age; however, the closer proximity of this 
family to the J5:2 permits some PA8-sized objects to enter the resonance. The flux from the Agnia family, however, is likely less than that of the Koronis family, as PA8-sized objects from the Agnia family are only just 
beginning to enter the resonance. Interestingly, the osculating inclination of 2007 PA8 ($\sim$ 2$^\mathrm{o}$) is most similar to the proper inclination of the Koronis family (see Figure \ref{f:ia}). While inclination is 
expected to vary moderately as a result of perturbations within the J5:2 \citep{2002Icar..156..399B}, the similarity in orbital parameters only strengthens the connection between 2007 PA8 and the Koronis family. 

Based on these results and those obtained by \citet{2014ApJ...791..120V} and \citet{2014A&A...567L...7N} it seems to be clear that there must be at least two source regions for 
the H chondrites, one in the inner part of the main belt, close to the 3:1 resonance, and one in the outer part, close to the 5:2 mean motion resonance with Jupiter. While these results 
represent a significant progress since the work of \citet{1998M&PS...33.1281G} there are still open questions regarding the origin of H chondrites. \citet{2013GeCoA.119..302M} 
noted that the bulk porosity of these meteorites ($\sim$ 10 \%) is consistent with typical values found for samples of high shock grade, suggesting a parent body with a "rubble-pile" 
structure for the H chondrites. However, they also noted that the high bulk density of (6) Hebe (the proposed parent body for these meteorites) is not compatible with a rubble-pile 
structure of pure H chondrite material. In addition, the identification by \citet{2014ApJ...791..120V} of several asteroids with H chondrite-like compositions close to the 3:1 resonance 
open the possibility that other objects in the inner part of the main belt could be the source of these meteorites. Thus, future studies should focus on determining physical properties such 
as bulk density and porosity for these other candidates. This could lead to the identification of an asteroid whose characteristics more closely resemble those measured in the laboratory 
for H chondrites.

Another important point that needs to be addressed is the contribution of each of these regions to the NEA population and, by extension, to the H chondrites that fall on Earth.
\citet{2013Icar..222..273D} used the dynamical model of \citet{2002Icar..156..399B} to determine the source region of the NEAs examined in their study. They found that NEAs with 
H chondrite compositions are more likely to be derived from the $\nu_{6}$ resonance, in the innermost region of the main belt. However, they noted that almost all NEAs with an H 
chondrite-like composition reside in a region near the 3:1 resonance. This apparent contradiction has yet to be explained. Furthermore, no contribution from the outer main belt was found. 
In light of the evidence presented in this work current dynamical models might need to be revised. This could help to better quantify the contribution of each region to the delivery of H 
chondrites to the near-Earth space.

\clearpage

\section{Summary}

We have obtained NIR spectra of the PHA (214869) 2007 PA8. Based on the measured spectral band parameters and the mineralogical analysis we found that the surface composition 
of this asteroid is consistent with that of H ordinary chondrites. In particular we determined that the olivine and pyroxene chemistries of this asteroid are Fa$_{18}$(Fo$_{82})$ and 
Fs$_{16}$, respectively, with an olivine abundance of 47\%. Moreover, the location of 2007 PA8 in the Band I center vs. BAR diagram, close to the H4 and H5 chondrites, and its low ol/(ol+px) 
ratio suggest that this object originated in the interior of a larger body that experienced certain degree of thermal metamorphism.

Based on their compositional affinity and proximity to the 5:2 mean motion resonance with Jupiter, three asteroid families (Agnia, Merxia and Koronis) were identified as the possible source region
for 2007 PA8. However, of these three families, Koronis seems to be the most likely source due to its older age and low proper inclination close to that of 2007 PA8.

\

{\bf{Acknowledgements}}

\

This research work was supported by NASA Near-Earth Object Observations Program grant NNX14AL06G (PI:Reddy). We thank the IRTF TAC for awarding time to this project, and to the IRTF TOs and MKSS staff for their 
support. The IRTF is operated by the University of Hawaii under Cooperative Agreement no. NCC 5-538 with the National Aeronautics and Space Administration, Office of Space Science, Planetary Astronomy Program. 
The authors would like to thank Dan A. Nedelcu for his review that helped to improve the manuscript.

\clearpage

\bibliographystyle{model2-names}

\bibliography{references}

\end{document}